\documentclass{aastex61}

\usepackage{amsmath,amssymb,amsfonts}
\usepackage{color}

\usepackage[T1]{fontenc}
\usepackage{txfonts}
\usepackage[]{graphicx}

\submitjournal{ApJ}

\shorttitle{Determining the nature of white dwarfs from low-frequency gws}
\shortauthors{Han \& Fan, 2017}

\begin{document}

\title{Determining the nature of white dwarfs from low-frequency gravitational waves}

\correspondingauthor{Xi-Long Fan}
\email{fanxilong@outlook.com}

\author{Wen-Biao Han}
\affil{ Shanghai Astronomical Observatory, Shanghai,  200030, People's Republic of China \\}
\affiliation{School of Astronomy and Space Science, University of Chinese Academy of Sciences,  Beijing 100049, China} 

\author{Xi-Long Fan}
\affiliation{Department of Physics and Mechanical and Electrical Engineering, Hubei University of Education, Wuhan 430205, China}




\begin{abstract}
An extreme-mass-ratio system composed of a white dwarf (WD) and a massive black hole can be observed by the low-frequency gravitational wave detectors, such as the Laser Interferometer Space Antenna (LISA). When the mass of the black hole is around $10^4 \sim 10^5 M_\odot$, the WD will be disrupted by the tidal interaction at the final inspiraling stage. The event position and time of the tidal disruption of the WD can be accurately determined by the gravitational wave signals. Such position and time depend upon the mass of the black hole and especially on the density of the WD. We present the theory by using LISA-like gravitational wave detectors, the mass-radius relation and then the equations of state of WDs could be strictly constrained (accuracy up to $0.1\%$). We also point out that LISA can accurately predict the disruption time of a WD, and forecast the electromagnetic follow-up of this tidal disruption event. 
\end{abstract} 




\section{Introduction} \label{sec:intro}

The era of gravitational wave (GW) astronomy arrived when the advanced Laser Interferometer Gravitational-Wave Obser-vatory (LIGO) observed the first gravitational wave event GW150914 \citep{gw15}. The latest observation of a double neutron star merger gives us a special chance to explore the universe using multi-message astronomy observations \citep{bn17a,bn17b}. The planning space-based gravitational wave detectors, such as Laser Interferometer Space Antenna (LISA),\footnote{https://www.lisamission.org} China's Taiji \citep{taiji} and Tianqin \citep{tianqin}, will provide an opportunity to observe extreme-mass-ratio inspirals (EMRIs) and intermediate-mass-ratio inspirals (IMRIs), which are composed by a stellar compact object and a massive black hole \citep{lisal3}. 

 White dwarf (WD) mass-radius relation,  which  is determined  by the physics of WDs,  is a fundamental  tool in modern astrophysics  \citep{tool,tool2}.  People usually believe that the theory of the equation of state (EoS) for WDs is clear. However, the varied compositions of WDs still yield different mass-radius relations \citet[see][and references inside]{book}. From Hipparcos data, the measurement accuracy of the mass-radius relation of WDs is very low ($\gtrsim 10\%$), and in pareicular some WDs fall onto the iron track -- a result that does not follow from standard stellar evolution theory (see Fig. 5.17 in \citep{book}). This situation asks for a more accurate measurement of the mass-radius relation of WDs, for determining the composition (C/O or Fe et. al.) and confirming the standard theory of WDs. 
 
 WDs inspiralling into MBHs (supermassive or intermediate massive), known as one type of EMRIs or IMRIs,  are potential GW sources for space-based gravitational wave detectors.  If the mass of black hole is less than  $10^6 M_\odot$, the WD will be very probably disrupted by the tidal force from the black hole before merger (i.e., before the innermost stable circular orbit, ISCO). Note that the burst of GWs generated by the disruption event \citep{rosswog07} and GWs emitted by a star colliding with an MBH \citep{east14}, are distinguishable from GWs in the inspiralling phase. Other signatures from tidal disruption of WDs by BHs, such as nucleosynthesis and the relativistic jet, have also been theoretically analyzed or numerically simulated in the literature (see \citep{rosswog09,macleod16,kawana17}, among others). Rate predictions for such events are extremely uncertain at present.  LISA likely detects several events during its lifetime\citep{sesana08}.    Since GW interferometers (like  LISA) are very sensitive to the phase of the signal, this phase difference is crucial for distinguishing the disruption position and time of the WD.  Such position and time are determined by the mass and radius of WD and the mass of the black hole.  LISA can estimate the masses of the WD and MBH with accuracies of $10^{-3}$ and $10^{-4}$ ,respectively, with SNR $>20$ \citep{lisal3}. Therefore,  by observing  the  GW waveform cutoff caused by the WD tidal disruption, one in principle can constrain the radius of WD accurately. Once the mass and radius of a WD are determined, then EoS or the composition of WD will be constrained rigidly too.  We point out that, with such kind of WD disruption events, LISA will constrain the mass-radius relation of WDs much better than current astronomical methods \citep{m-r,m-r2,m-r3}.      

\section{Method} \label{sec:intro}
 The tidal radius for a WD can be easily estimated. Ignoring the rotation of WD and taking it as a rigid body, the Roche limit is
\begin{align} \nonumber 
r_{\rm tidal} &= \sqrt[3]{2} R_{\rm WD} \left(\frac{M_{\rm BH}}{m_{\rm WD}}\right)^{1/3} \\
                    &= \sqrt[3]{2} M_{\rm BH} \left(\frac{\rho_{\rm BH}}{\rho_{\rm WD}}\right)^{1/3} \,, \label{rtidal}
\end{align}
where $R_{\rm WD}$ is the radius of WD. We find that the tidal radius is independent of the spin of the black hole (BH), but relates with the density of WD and mass of BH. A tidal disruption event (TDE) will happen at the position of tidal radius. Due to the TDE, the GW signals of this EMRI event disappears once the WD enter the tidal radius. Therefore, the GW frequency at this moment is the cutoff frequency, which can be calculated easily. At the radii of tidal disruption, the orbital frequency is
\begin{align}
\tilde{\Omega}_{\rm tidal} = \frac{1}{\tilde{r}_{\rm tidal}^{3/2}+q} \,, \label{omega}
\end{align}
where the variables with tildes mean dimensionless, and $q$ is the dimensionless spin parameter of the black hole defined as $q \equiv J/M^2$ ($J$ is the rotating angular momentum). Due to the tidal disruption, there is a cutoff frequency of GWs for observation. From Eq. (\ref{omega}), changing to the SI units, the cutoff frequency of the dominant (2, 2) mode is 
\begin{align}
f^{\rm c}_{22} =0.64625\left[\left(\frac{r_{\rm tital}}{1.47663\times10^5 {\rm km}}\frac{10^5 m_\odot}{M}\right)^{3/2}+q\right]^{-1} \left(\frac{10^5 m_\odot}{M}\right) {\rm Hz} \,, \label{cutoff}
\end{align}
or 
\begin{align}
r_{\rm tidal} = 1.47663\times10^5 \left[\left(\frac{f^{\rm c}_{22}}{0.64625}\frac{M}{10^5 m_\odot}\right)^{-1}+q\right]^{2/3} \left(\frac{M}{10^5 m_\odot}\right)  {\rm km} \,. \label{rtidal2}
\end{align}
The $r_{\rm tidal}$ is decided by the cutoff frequency, the mass and spin of MBH. It is well known that LISA can measure the masses of the binary system and spin of MBH precisely from the insprialling waveforms \citep{lisal3}.  By using the TDE, in principle, one can independently obtain the tidal radius of WDs from the GW's cutoff frequency. The error of $r_{\rm tidal}$ depends on the measurement accuracy of $M, ~q$ and frequency resolution. Once we get the value of $r_{\rm tidal}$, with Eq. (\ref{rtidal}), the radius of WD can be obtained at the same accuracy of the tidal radius,
\begin{align}
R_{\rm WD} &= r_{\rm tidal} \left(\frac{1}{2}\frac{M}{m_{\rm WD}}\right)^{-1/3} \label{rwd2} \,. 
\end{align}

However, since we do not know the frequency resolution of LISA data, we replace it with a waveform dephasing resolution. LISA is very sensitive to the phase of gravitational wave signals.  Using matched filtering, LISA will be able to determine the phase of an EMRI to an accuracy of half a cycle \citep{babak14}. Correspondingly, we assume one-cycle (2$\pi$) waveform dephasing will be recognized by LISA, and one-cycle waveform dephasing just needs $\delta t = 1/{f}_{22}^{\rm c}$ s. 

We have \citep{hughes00}
\begin{align}
\dot{r} = -\frac{c_{21}}{d}\dot{E}-\frac{c_{22}}{d}\dot{L_z} \,, \label{rdot}
\end{align} 
where the coefficients are listed as
\begin{align} \nonumber
c_{21} \equiv \,& 2Er^{5}-6EMr^{4}+4a^{2}Er^{3}+2a(L_{z}-2aE)Mr^{2}+2a^{4}Er-2a^{3}(L_{z}-aE)M\,, \\ \nonumber
c_{22} \equiv \,&2aEMr^{2}-2a^{2}L_{z}r+2a^{2}(L_{z}-aE)M , \, \\ \nonumber
d \equiv \, &-2(1-E^{2})Mr^{4}+8(1-E^{2})M^{2}r^{3}+[Q+L_{z}^{2}-5a^{2}(1-E^{2})- \\ \nonumber
& \,6M^{2}]Mr^{2}+2[a^{2}(3-E^{2})+2aEL_{z}-(L_{z}^{2}+Q)]M^{2}r+2(L_{z}^{2}+Q)M^{3}- \\ \nonumber
& \,4aEL_{z}M^{3}+a^{2}(2E^{2}M^{2}-L_{z}^{2}-Q)M-a^{4}(1-E^{2})M \,,
\end{align} 
where $E, ~L_z$, and $Q$ are energy, angular momentum, and Carter constant respectively. If we constrain the particle on the equatorial plane of the Kerr black hole, then $Q=0$. $E, ~L_z$ can be analytical obtained for the circular orbit cases. The gravitational fluxes $\dot{E}$ and $\dot{L_z}$ can be calculated very accurately by solving the Teukolsky equations in frequency domain \citep{han10,han11}.  From Eq. (\ref{rdot}), one can calculate the uncertainty of radii $r$ with $\delta t$ allowed for one-cycle dephasing and $r_{\rm tidal}$ from Eq.  (\ref{rtidal2}). 

An analytical approximation for the mass-radius relation for nonrotating WDs \citep{nauenberg72} is
\begin{align}
\frac{R_{\rm WD}}{R_\odot} = \frac{0.0225}{\mu}\frac{[1-(m_{\rm WD}/m_{\rm max})^{4/3}]^{1/2}}{(m_{\rm WD}/m_{\rm max})^{1/3}}\,, \label{rwd} 
\end{align}
where $\mu$ is the mean molecular weight. It is usually set equal to 2, corresponding to helium and heavier elements, which is appropriate for most astrophysical WDs. The Chandrasekhar's limit of mass of WD is  $m_{\rm max} = 5.816m_\odot/\mu^2$.

However, as explained in Nauenberg's paper, Eq. (\ref{rwd}) only takes into account electron degeneracy. This induces an uncertainty on the order of 3\%. And WD binary system simulations have shown us that WDs heat up a lot due to tidal effects, making Eq. (\ref{rwd}) even less accurate. Thus, with the measured $R_{\rm WD}$ by GWs from Eq. (\ref{rwd2}), we can validate the accuracy of Eq. (\ref{rwd}) or, alternatively, determine the value of $\mu$.

In addition, with Eq. (\ref{rwd}), one can predict the TDE time by using the mass and spin parameters from the insprialling waveform. At a given GW frequency, the total remaining time until the end of the inspiral is \cite{finn00}
\begin{align}
T_{\rm rem} = \frac{1.41\times10^6 {\rm sec}}{(f_{\rm GW}/0.01 {\rm Hz})^{8/3}} \left(\frac{10 m_\odot}{m_{\rm WD}}\right)\left(\frac{10^6 m_\odot}{M}\right)^{2/3} \cal{T} \label{tremain} \,, 
\end{align}
where $\cal{T}$ is the relativistic correction, which can be calculated with accurate Teukolsky-based fluxes. From the radius-mass relation (\ref{rwd}) and Eq. (\ref{rtidal}), we can determine the tidal radius and calculate the remaining time of TDE to ISCO.  At any given $r > r_{\rm tidal}$, one can calculate the remaining time to ISCO. The difference of two remaining times is just the expected time of TDE relative to the position at radius $r$. 
              
\section{WD's EoS constraint results} \label{sec:res}
\begin{figure}[!h]
\begin{center}
\includegraphics[height=4.0in]{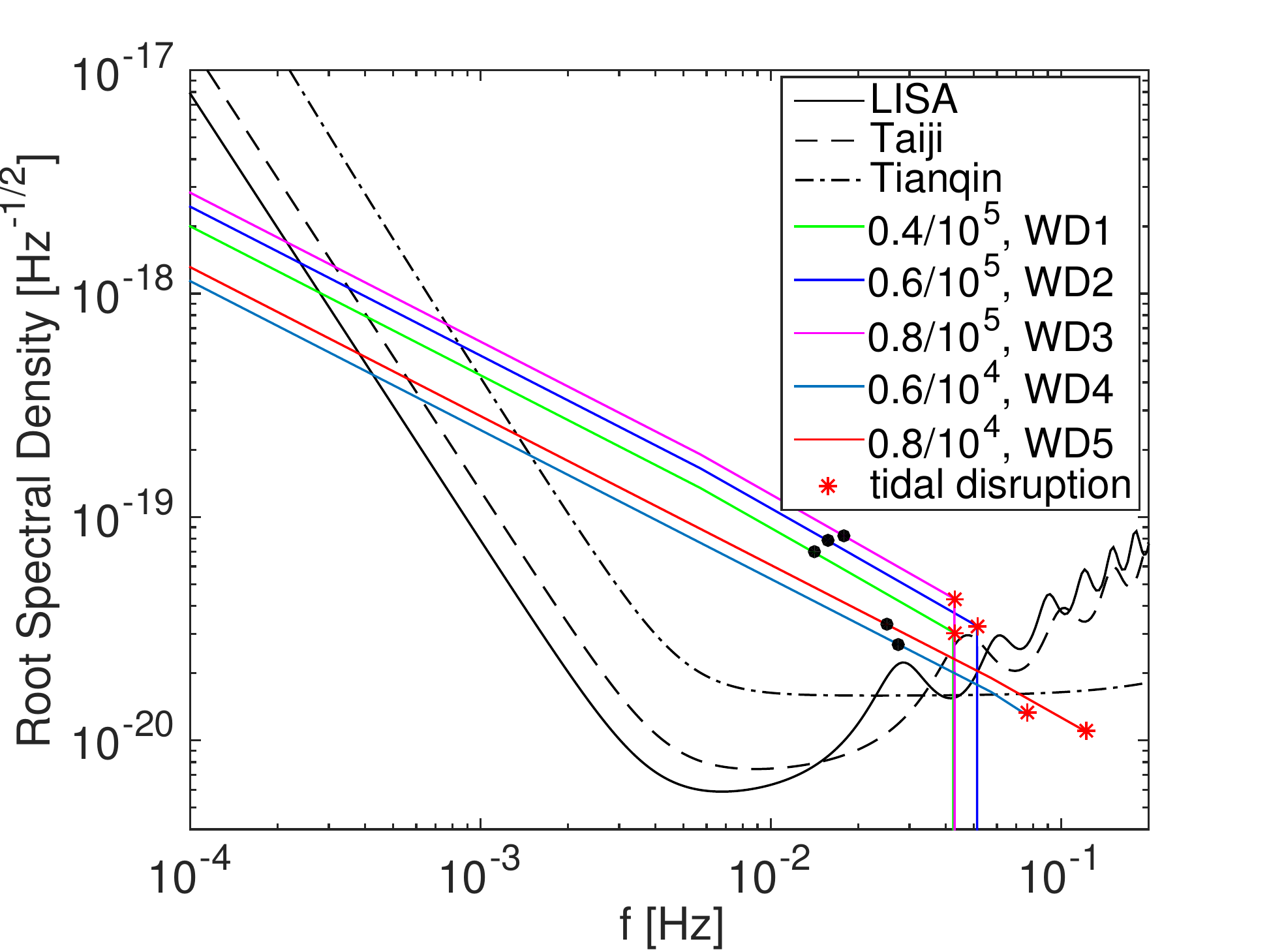}
\caption{Noise curves of LISA, Taiji, and Tianqin; the root of the power-spectrum-density (PSD) curves of GWs from five WD tidal disruption events. All these five disruption events happen before the WDs arrive at ISCO, and three of the TDEs are in the sensitive band of detectors. These three TDEs take place, respectively 21, 16, and 42 days (corresponding 0.4, 0.6, and 0.8 solar mass WDs and $10^5$ solar mass MBH, the dimensionless spin parameter of the MBH is $q = 0.9$.) before they arrive at ISCO. Two of the five TDEs happen out of the sensitive band ( WD4 and WD5, densities are 4.2525 $\times 10^8$ and  1.0712 $\times 10^9$ kg/m$^3$ respectively). The black points represent the time of 1 year before the tidal disruption events. }  \label{simple1}
\end{center}
\end{figure}

In Fig. \ref{simple1}, we consider five WDs with different masses and densities. Three of them (WD1: 0.4 $m_\odot$ \& $\rho = 1.4648\times 10^8$ kg/m$^{3}$, WD2: 0.6 $m_\odot$ \& $\rho = 2.1512\times 10^8$ kg/m$^3$, and WD3: 0.8 $m_\odot$ \& $\rho = 1.4714\times 10^8$ kg/m$^3$) are inspiralling into an MBH with mass $10^5~m_\odot$ and spin $q = 0.9$. The tidal disruption turns up at $r_{\rm tidal}  = 5.8612, 5.1565$， and $5.8524$$M_{\rm BH}$ respectively. The radii of ISCO  is 2.32 $M_{\rm BH}$, so all the three WDs will be disrupted before merger and inside the sensitive band of GW detectors. 

The cutoff point (*) in Fig. \ref{simple1} represents the disruption event, and the GW frequency at the moment of TDE is just the cutoff frequency. In this paper, we request $f^{\rm c}_{22} < f^{\rm isco}_{22}$ so that the WD can be tidally disrupted by the black hole. If using this relation and Eq. (\ref{rwd}) to Eq. (\ref{rtidal}), we have
\begin{align}
r_{\rm tidal} &= 0.05098 \mu^{-5/3} \left[1-\left(\frac{\mu^2}{5.816} \frac{m_{\rm WD}}{m_\odot}\right)^{4/3}\right]^{1/2}\left(\frac{m_\odot}{m_{\rm WD}}\right)^{1/3}\left(\frac{M_{\rm BH}}{m_{\rm WD}}\right)^{1/3} R_\odot \,. \label{rtmu} 
                    \end{align}
Applying this relation to Eq. (\ref{cutoff}), we find that the cutoff frequency is not sensitive to the mass of the black hole, but very sensitive to the mass of the WD and the parameter $\mu$ (or the density). The cutoff frequencies of the first three systems in Fig. \ref{simple1} are 0.0428 Hz, 0.0513 Hz, and 0.0429 Hz for the WD1, WD2, and WD3, respectively, and these three TDEs happen inside the sensitive band of detectors.

The phase of the EMRI waveform will stop evolving at the moment of TDE. Thus the value of the phase at the cutoff frequency is decided by the moment of TDE or tidal radius. From Eq. (\ref{rtidal}), the tidal radius is decided by the masses of the binary and the radius of WD. From the inspiralling GWs, LISA can accurately determine the masses. Therefore, the phase of the cutoff waveform decides the radius of a given WD based on our analysis. The measurement accuracy of the waveform phase by LISA can be a fraction of $2\pi$ during $10^5$ cycles, i.e.,  a fractional phase accuracy of up to $10^6$. All information about the EMRI is encoded in the GW phase and thus we can expect to make measurements of the intrinsic parameters to this same fractional accuracy \citep{babak14}. Here, even if we assume a $2\pi$ phase accuracy of the waveform,  we will see that the radius or density of a WD can be totally constrained from the observation of GWs of a disruption phenomenon. 

At $r_{\rm tidal}  = 5.8612, 5.1565$, and $5.8524$ $M_{\rm BH}$, the energy fluxes are $6.2603 \times 10^ {-4}, ~1.1286 \times 10^ {-3}$, and $6.3040 \times 10^{-4} (m_{\rm WD}/M_{\rm BH})^2$ respectively, for the mentioned three WDs. From Eq. (\ref{rdot}), $\dot{r} = -2.0093\times10^{-7}, ~-4.3275\times10^{-7}$ and $-4.0357\times10^{-7}$ at the moment of tidal disruption. If LISA can measure the phase of GWs in a precision of one cycle ($2\pi$), then the $\delta t =1/f_{22}^{\rm c} \approx 47.4,  39.6, {\rm and} ~ 47.3 M_{\rm BH}$ (23.7, 19.8, and 23.7 s for a $10^5$ solar MBH) for the above three WDs respectively. In a result, the tidal disruption position can be determined in a relative uncertainty $\delta r_{\rm tidal}/r_{\rm tidal}$ only $1.6252 \times 10^{-6}, ~3.3245 \times 10^{-6}$ and $3.2621 \times 10^{-6}$ respectively, for the above three cases. From Eq. (\ref{rtidal}), it means LISA can determine the radii of WD with the same relative uncertainty if we can totally determine the masses of the WD and the black hole. Equivalently, the density of WDs can be constrained in relative uncertainties  $4.8756 \times 10^{-6}, ~9.9735 \times 10^{-6}$ and $9.7863 \times 10^{-6}$ respectively.  

From the masses of the system, and a theoretical model of WD's mass-radius relation such as in Eq. (\ref{rtmu}), we can predict the tidal disruption time by Eq. (\ref{tremain}). In Fig. \ref{simple1}, we plot a black dot to represent the moment of 1 year before TDE. A small error (0.1\%) of the radius will induce an incorrect prediction time of about a few hours.  

In practice, one cannot know the exact spin and mass of a system. The measurement uncertainties of the mass of a WD and the mass and spin of a black hole will also contribute to the measurement error of the tidal disruption position. LISA can estimate the masses of the WD and MBH at an accuracy of $10^{-3}$ and $10^{-4}$ respectively, and the spin of the BH at a level of $10^{-3}$ with SNR $>20$ \citep{lisal3}.  From Eq. (\ref{rtidal}), one can immediately determine that the error of the black hole's mass will induce an error of estimation $\delta R_{\rm WD}/R_{\rm WD} = 1/3 \delta M_{\rm BH}/M_{\rm BH} \sim 3.3\times10^{-5}$. At the same time, the error of spin of the BH will induce  $\delta R_{\rm WD}/R_{\rm WD} < 2/3 \delta q/q < 10^{-3}$ and $\delta m_{\rm WD}/m_{\rm WD}$ will produce an error of WD radius $\sim 3.3\times10^{-4}$. In this way, the constraint accuracy of the WD radius is mainly influenced by the precision of the WD mass and the BH's spin. Despite this, while observing a WD tidal disruption with LISA, the mass-radius ratio of a WD can still be constrained with an accuracy $ \sim 0.1\%$ level, which will be much better than the current results from astrophysical observations \citep{m-r,m-r2,m-r3}(around 10\% level). A few of these types of events will totally decide the composition and EoS of WDs. 

\begin{table}[!h]
\centering
  \caption{The ideal constraint results of WD's radius. The third column is the tidal radius for different WDs, and the forth one is the TDE time before the compact object arrives at ISCO. The last column is the constraint accuracy of the radius with one-cycle-dephasing measurement precision. (*) means that the estimation is obtained by assuming that the other parameters, such as mass and spin of the system, are known exactly. The mass of black hole is $10^5$ solar masses and spin $q = 0.9$.} \label{table}
  \begin{tabular}{c|c|c|c|c|c|c}
  \hline\hline
   WD  & mass & density [kg/m$^3$] & $r_{\rm td}$ & $T_{\rm td}$ & $f^{\rm c}_{22}$ &  $|\delta R_{\rm WD}|/R_{\rm WD}$ (*)  \\
  \hline
  WD1& $0.4 m_\odot$ &$1.4648\times 10^8$   & 5.8612$M_{\rm BH}$& 42 day & 0.0428 Hz & $1.6252 \times 10^{-6} $ \\
  WD2& $0.6 m_\odot$ & $2.1512\times 10^8$  & 5.1565$M_{\rm BH}$& 16 day & 0.0513 Hz &  $3.3245 \times 10^{-6} $ \\
  WD3& $0.8 m_\odot$ & $1.4714\times 10^8$  & 5.8524$M_{\rm BH}$& 21 day & 0.0429 Hz &  $3.2621 \times 10^{-6} $ \\
  \hline\hline
  \end{tabular}
\end{table}

Not all WD-MBH systems can be used to constrain the equations of state of WDs by TDEs with only GW signals. If the mass of the black hole is too large (something like a million solar masses), WDs will survive until merger. If the mass is too small ($ < 10^4 ~m_{\odot}$), WDs will be disrupted with a high cutoff frequency, which is not in the sensitive band of LISA. Two EMRIs (0.6 and 0.8 solar massive WDs with a $10^4 m_\odot$ BH) in Fig. \ref{simple1} show that their TDEs are out of the sensitive band of LISA. In this way, we cannot use only gravitational waves to constrain the WD's mass-radius relation. However, the GW signal will tell us the exact moment of TDE based on a given mass-radius relation of WD. Therefore, by using a multi-message astronomical observations, i.e., a GW observation of the inspiral waveforms and electronic-magnetic (EM) observation of TDEs will also offer an opportunity to constrain the EoS of WDs. We can estimate that a 0.1\% WD radii difference will induce 0.57 and 2.59hr TDE's EM signals arrival difference for the above two EMRIs.

\begin{figure}[!h]
\begin{center}
\includegraphics[height=4.0in]{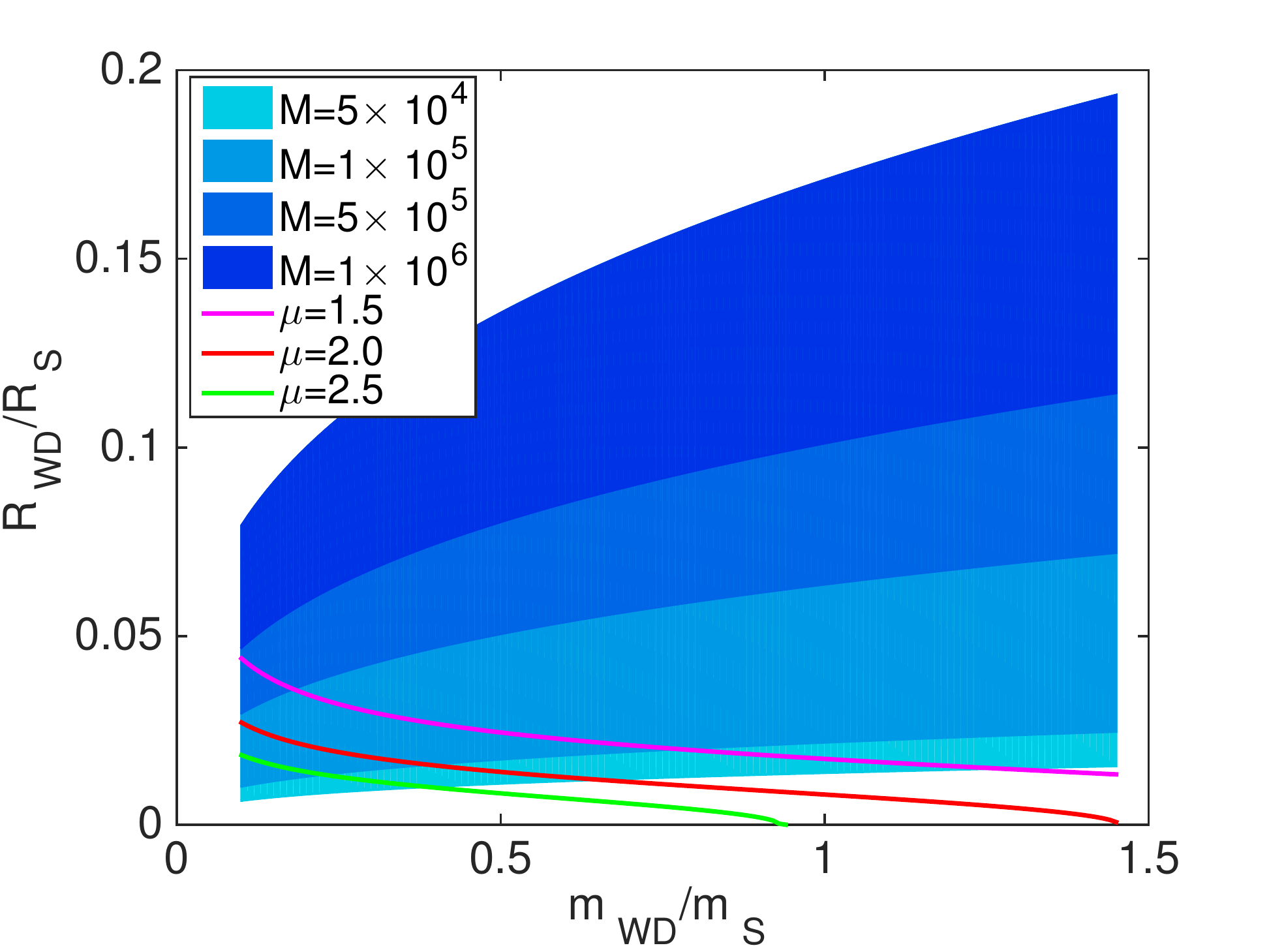}
\caption{Different colored shadows represent the range of masses and radii of compact objects with tidal disruption events seen by LISA. The larger the BH mass is, the smaller the range of mass-radius seen by LISA will be. The solid lines represent the mass-radius relations of astrophysical WDs with different compositions. }  \label{simple2}
\end{center}
\end{figure}

In Fig. \ref{simple2}, we plot the range of the mass-radius relation of arbitrary compact objects with TDEs, which can be observed by LISA, and the mass-radius relation of different kinds of WDs. For a typical WD ($\mu = 2$), we find that  LISA can observe the tidal disruption of a small mass WD ($\lesssim 0.5 m_\odot$).

\section{Discussion} \label{sec:res}
 In this paper, we assume a WD-MBH system with appropriate masses, i.e., a low massive WD and a $10^4 \sim 10^5 m_\odot$ black hole. This makes sure the tidal disruption happens inside the LISA band. By observing the waveforms and its cutoff frequencies of such kind of EMRIs, we can independently obtain the radii of WDs. Our method is independent of any WD models. Our analysis shows that  the uncertainty of a WD's radius with GWs can be as good as 0.1\%. Therefore, we can constrain the mass-radius relation (equivalently, EoS) of WDs in a very high precision. The mass-radius relation obtained from GWs can be used to validate the theoretical models of WDs. In addition, one can use GWs and the theoretical WD model to predict the TDE time. This makes sure that the telescopes can be ready  in advance to monitor the electromagnetic counterpart due to the TDE. The comparison of the real TDE time and the predicted one can also be used to constrain the EoS of WDs.
 
In this way, it is interesting that even if a TDE happens outside of the sensitive band of LISA, we can also predict the TDE time based on the GW observations. when the mass of a WD is a little large ($\approx 1 m_\odot$), or the mass of a BH is around $10^3$, the TDE will happen with a higher cut-off frequency, which is outside of the LISA's band. For example, for the $0.6/10^4 m_\odot$ and $0.8/10^4 m_\odot$ WD-MBH systems, though the TDEs happen outside of the sensitive band of LISA, GW observation can still predict the TDE time very accurately.  Assuming the TDE is finally observed by telescopes at the prediction time and coincides with the position and  distance of the GW source, it will also be very meaningful for the constraint of the mass-radius relation of WDs. 

\section*{Acknowledgement}
This work is supported by the National Natural Science Foundation of China, No. 11773059, No. 11673008, No. U1431120, and No. 11690023, and by the Key Research Program of Frontier Sciences, CAS, No. QYZDB-SSW-SYS016. W.H. is also supported by the Youth Innovation. We thank the anonymous referees for valuable comments and suggestions that helped us to improve the manuscript.



\end{document}